\documentclass[10pt]{article}
\usepackage{amsmath,amssymb}
\batchmode

\newtheorem{defn}{Definition}
\newtheorem{lem}{Lemma}
\newtheorem{thm}{Theorem}

\newcommand{\pr}{\noindent{\bf Proof}. }

\newcommand{\res}{\noindent{\bf Remarks}. }

\newcommand{\al}{\alpha}
\newcommand{\De}{\Delta}
\newcommand{\de}{\delta}

\newcommand{\Ga}{\Gamma}
\newcommand{\ka}{\kappa}
\newcommand{\La}{\Lambda}

\newcommand{\Om}{\Omega}
\newcommand{\om}{\omega}

\newcommand{\ep}{\epsilon}

\newcommand{\cB}{{\cal B}}
\newcommand{\cC}{{\cal C}}
\newcommand{\cD}{{\cal D}}

\newcommand{\cO}{{\cal O}}

\newcommand{\cH}{{\cal H}}
\newcommand{\cS}{{\cal S}}

\newcommand{\cF}{{\cal F}}

\newcommand{\cL}{{\cal L}}

\newcommand{\bbR}{{\mathbb{R}}}

\newcommand{\bbC}{{\mathbb{C}}}

\begin{document}
  
\title{Multi-particle Schrodinger operators with point interactions in the plane}
\author{ 
J. Dimock\\
Dept. of Mathematics \\
SUNY at Buffalo \\
Buffalo, NY 14260 \\
\and
S.G. Rajeev\\
Dept. of Physics and Astronomy\\
University of Rochester\\
Rochester, NY, 14627\\}

\maketitle

\begin{abstract}
We study a system of N- bosons in the plane interacting with 
delta function potentials.  After a coupling constant renormalization
we show that the Hamiltonian defines a self-adjoint operator and obtain
a lower bound for the energy.  The same results hold if one includes
a regular inter-particle potential.
\end{abstract}

\section{Introduction}

We consider  $N-$bosons of mass $m$ in  $\bbR^2$ interacting with delta function 
potentials of strength  $g$.   The Hamiltonian for the system is 
\begin{equation}
H  =  \sum_{i=1}^N  \frac{-\De}{2m}\ \   -  \sum_{1\leq i < j \leq N}  g \de (x_i -x_j)
\end{equation}
defined on $\cH_N$ = the $N$-fold symmetric tensor product of  $\cL^2(\bbR^2)$ with itself.
The problem is  to make sense of this  as a self adjoint operator.   This is necessary 
in order that the global  dynamics    $\psi_t = e^{-iHt}\psi_0$ be well-defined. However   the
expression is quite a singular  and one finds that the coupling constant $g$ must be
renormalized to have a chance of success. 

The problem is  
fairly well understood for $N=2$. We give a treatment below which involves
introducing  a momentum cutoff, choosing a cutoff dependent coupling constant, and  then showing
that as the cutoff is removed the Hamiltonians have a self-adjoint limit in the sense
of resolvent convergence.

The $N=2$ problem has also been considered by 
Albeverio, Gesztesy, Hoegh-Krohn,
and Holden  \cite{AGHH88}.  They take a different 
approach which involves specifying boundary conditions when the points coincide.
We show that our    results are equivalent to theirs.

Our main interest is in   general $N$ and the challenge 
is to incorporate the wisdom gained for the two particle case into the multi-particle 
setting.   Our solution involves 
introducing  fictitious particles known as
\textit{angels} which serve as markers for two-particle subsystems.  This approach
 was developed by 
one of us in the papers  \cite{HJ98}, \cite{Raj99}.  The present paper is a 
rigorous version of this work.   The  main result is  again   a proof that 
the cutoff renormalized operators have a self-adjoint limit in the
sense of resolvent convergence.
  We also obtain a lower bound for the Hamiltonian.

This problem was previously  considered by Dell'Antonio, Figari,
Teta \cite{DFT94}, who also consider  $d=3$. Our  results for $d=2$ are very similar to theirs. 
However the proofs are  rather different. They  use a concept of $\Ga$-convergence rather
than resolvent  convergence.  Also their `bare   coupling constant' 
depends on  $N$ and momentum as well as the  cutoff, whereas ours depends only on the
cutoff.

The present work seems to have some advantages in simplicity and flexibility.  As 
evidence of this we  obtain the new result that essentially  the same conclusions  hold if we   include a regular
inter-particle  potential in addition to  the delta function.

\section{Two particles}

\subsection{point interaction}

We begin with a discussion of the case  $N=2$.   Taking mass $m=1$ and 
passing to to center of mass coordinates we have the Hamiltonian  
\begin{equation}  \label{config}
  H  = -\De - g \de 
\end{equation}
on the space  $\cL^2(\bbR^2)$.
In momentum space
\begin{equation}  \label{momentum}
 (H\psi)(p)  = p^2 \psi(p)  -   \frac{g}{(2\pi)^2}\int \psi(q) dq 
\end{equation}
This  operator does not map into $\cL^2(\bbR^2)$ 
and cannot determine a dynamics as such. 

Instead we consider approximate Hamiltonians 
\begin{equation}
 (H_{\La}\psi)(p)  = p^2 \psi(p)
 -   \frac{g_{\La}}{(2\pi)^2}\ \rho_{\La}(p) \int \rho_{\La}(q)\psi(q) dq
 \end{equation}
where  $\rho_{\La}$  is the characteristic function of  $|p| \leq  \La$.
We define
\begin{equation}   P_{f}\psi =  f(f, \psi)\end{equation}
(If  $\|f\| =1$ this is the projection onto $f$.) 
Then we can write with  $H_0=p^2$
\begin{equation} 
 H_{\La}  = H_0 -   \frac{g_{\La}}{(2\pi)^2}\ P_{\rho_{\La}} 
 \end{equation}
This is a bounded perturbation of the self-adjoint operator $H_0$ 
and  and so is   self-adjoint on  $D(H_0)$ 
(Kato's theorem \cite{Kat66}).  We define the resolvents
\begin{equation}
R_0(E)  = (H_0-E)^{-1}\ \ \ \ \ \ \  R_{\La}(E)  = (H_{\La}-E)^{-1}
\end{equation}
when they exist. If they exist as bounded operators one says that
$E$ is in the resolvent set of the operator.  The resolvent set for 
$H_0$ is  $  \bbC - [0,\infty)$. 
 Since the perturbation is  rank one the resolvent $ R_{\La}(E)$ can be explicitly calculated.  For 
$E \in   \bbC - [0,\infty)$ one finds that   $E$ is in the resolvent set for 
 $H_{\La} -E$  if and only 
if   
\begin{equation}  \label{point}
 (2 \pi)^2 g_{\La}^{-1} \neq 
(\rho_{\La}, R_0(E)  \rho_{\La})
\end{equation}
 in which case  
\begin{equation}
 R_{\La}(E) 
  =  R_0(E) 
 + \left(\frac {1}{  (2 \pi)^2 g_{\La}^{-1} - 
(\rho_{\La}, R_0(E)  \rho_{\La})}
\right) P_{ R_0(E) \rho_{\La}} 
\end{equation}
Indeed if  (\ref{point}) holds then an explicit calculation shows that
the right side provides a bounded inverse for $H_{\La} -E$.  On the other hand if   
$(2 \pi)^2 g_{\La}^{-1} = (\rho_{\La}, R_0(E)  \rho_{\La})$
then 
\begin{equation}
(H_{\La}-E)R_0(E) \rho_{\La} =
  \left( 1-   \frac{g_{\La}}{(2\pi)^2} (\rho_{\La}, R_0(E)  \rho_{\La})
\right) \rho_{\La}  =0
\end{equation}
and $R_0(E) \rho_{\La} \neq 0$  so $E$ is an eigenvalue of  $H_{\La} -E$ and not in the resolvent
set.

Now we introduce a new parameter  $\mu >0$ and make the choice
\begin{equation}  \label{g}
   g_{\La} =  g _{\La}(\mu) 
 =(2\pi)^2 \left( \int_{ |p| \leq  \La}  (p^2+ \mu^2)^{-1} dp  \right)^{-1}
  \end{equation}
Thus $ g_{\La}$ goes to zero logarithmically  as  $\La \to \infty$.
Then we can write
\begin{equation}  R_{\La}(E)=      R_0(E) 
 +  \xi_{\La}(\mu^2,-E)^{-1} \ P_{ R_0(E) \rho_{\La}}
\end{equation}
where  
\begin{equation}  
\begin{split}
  \xi_{\La}(a,b)  \equiv &  \int_{ |p| \leq  \La}  (p^2+ a)^{-1} dp  
\ - \   \int_{ |p| \leq  \La}  (p^2 +b)^{-1} dp  \\
\end{split}
\end{equation}
For $a,b >0$ we have  
\begin{equation}   \label{cancel}
\begin{split}
  \xi_{\La}(a,b)  \equiv &  \pi  \log( \frac{\La^2}{a}  +1  )  -  \pi   \log( \frac{\La^2}{b} +1 )
\\
 = & \pi  \log( \frac{1}{a}  + \frac{1}{\La^2}  )  -  \pi   \log( \frac{1}{b} +
\frac{1}{\La^2}) \\
\end{split}
\end{equation}
In the last step we have canceled the divergence in each term by adding and subtracting  $\pi \log
\La^2$. Now it is a simple matter to take the limit  $\La \to \infty$ and get 
\begin{equation}
\xi(a,b) =  \pi  \log( b/a )
\end{equation}

\begin{thm} \ \   \label{one}
\begin{enumerate} 
\item For $E$ real and not in $ \{-\mu^2\}\cup [0,\infty) $ 
the strong limit   $R(E) = \lim_{\La \to\infty} R_{\La}(E)$ exists  
and is given by  
\begin{equation}   \label{resolvent}
 R(E) =  R_0(E)  +  \xi(\mu^2,-E)^{-1} \ P_{\Om_E}
\end{equation}
where    $\Om_E  \in \cL^2(\bbR^2)$  is defined by 
\begin{equation}
\Om_E(p) =  (p^2-E)^{-1}
\end{equation}
\item   $R(E)$ is invertible
\item  For $E$ complex and  not in   $ \{-\mu^2\}\cup [0,\infty)$ the limit   $R(E) = \lim_{\La \to
\infty} R_{\La}(E)$ exists. 
There is a self-adjoint operator $H(\mu)$ such that   $R(E)  =  (H(\mu) -E)^{-1}$.
\end{enumerate}
\end{thm}
\bigskip

\pr
\begin{enumerate}
\item  Under our hypotheses   $\xi(\mu^2,-E) =  \pi \log(-E/\mu^2) \neq  0 $.
Hence     $\xi_{\La}(-E,\mu^2) \neq  0$
for    $\La $ sufficiently large and 
$\xi(\mu^2,-E)^{-1} = \lim_{\La \to \infty}  \xi_{\La}(\mu^2,-E)^{-1}$.
We also have in  $\cL^2(\bbR^2)$ the limit 
$  \Om_E    =  \lim_{\La \to \infty}  R_0(E) \rho_{\La}$
The result follows.

\item 
To show  the null space of 
$R(E)$ is  $\{0\}$ it is sufficient to   find a dense set  $\cD  \subset D(H_{\La})$ such
that  for   $\eta \in \cD$ we have  the existence of 
 $\eta^*  =  lim_{\La \to \infty} (H_{\La}-E) \eta $.   For then 
if   $R(E) \psi =0$
we have 
\begin{equation}
(\eta, \psi)  =  lim_{\La \to \infty} ((H_{\La}-E) \eta,  R_{\La}(E)  \psi) 
=  ( \eta^*, R(E) \psi) =0
\end{equation}
for all $\eta \in  \cD$ and hence  $\psi =0$ 

For our domain  $\cD$ we pick $u \in   \cS(\bbR^2)$ so the  the Fourier transform $\hat u$  
is in   $\cC_0^{\infty} (\bbR^2 -\{0\})$. 
For $u$  in this domain  we have 
\begin{equation}
 (H_{\La} u )(p)  = p^2  u(p)
 -   \frac{g_{\La}}{(2\pi)^2}\ \rho_{\La}(p) \int \rho_{\La}(q) u (q) dq
 \end{equation}
We argue that the second term converges to zero
 so that $H_{\La} u  \to  H_0 u$. 
Since $g_{\La} 
\to 0$ and $\| \rho_{\La}  \|  =  \sqrt{\pi} \La$ it suffices that 
 $\int \rho_{\La}(q) u (q) dq  = \cO(\La^{-1})$. 

To see this first
 replace  $\rho_{\La}(q)  = \rho_1(q/\La) $ by 
 $\rho^*_{\La}(q)  = \rho^*(q/\La) $  where  $\rho^*$ is smooth approximation
to the characteristic function of the unit disc.  The difference is  $ \cO(\La^{-n})$
 for any $n$, and so it suffices to show  $ \int \rho^*_{\La}(q) u (q) dq  = \cO(\La^{-1})$.

 Since    $\hat  u  \in  \cC_0^{\infty} (\bbR^2 - \{ 0 \} )$  we have     $\hat  v(x) = |x|^{-2} \hat  u(x)$  in the
same  space 
and so  $ u  =  - \De_q  v$  for some   $v \in \cS(\bbR^2)$.
 Then after integrating 
by parts  
\begin{equation}
\int \rho^*_{\La}(q) u (q) dq  = 
\int ( - \De_q   \rho^*_{\La})(q)  v (q) dq  
\end{equation}
This is  $ \cO ( \La^{-2})$
since   $ |\De_q   \rho^*_{\La}(q) |$ is   $\cO ( \La^{-2})$ and  $ v (q) $ is 
rapidly decreasing.

\item   This follows from the first two parts and a version of the Trotter-Kato 
theorem quoted in the Appendix.

\end{enumerate}

\bigskip

\res
\begin{enumerate}

\item  
 The resolvent has   a simple pole at   $E =  - \mu^2$ so  $H(\mu)$ has 
the  eigenvalue  $ -\mu^2$.   The residue is the projection onto 
the eigenspace which we see is spanned by   $\Om_{-\mu^2}(p)  = (p^2 + \mu^2)^{-1}$.
This is  a bound state.

\item   
Our approach to this problem follows a path well-known to physicists.
The problem is usually  cited  as an example of \textit{dimensional transmutation}
in which a  model without a length scale   (the coupling constant  $g$
is dimensionless)
upon  renormalization gains a length scale   (namely $\mu^{-1}$) 
\cite{Hua82}. This phenomenon is expected to occur in gauge theories in four dimensions.

\item  Let us compare our result with the result of  Albeverio et. al.  \cite{AGHH88}.
 They consider  $-\De$ on 
on  $\cL^2(\bbR^2 \backslash \{0\})$ and then obtain various self-adjoint extensions
by imposing boundary conditions at the origin.   They obtain 
a family of self-adjoint operators
indexed by a parameter $ \alpha $  taking all real values.
They also have an explicit formula for 
the resolvent  (a Krein formula)  which is just like
our equation  (\ref{resolvent})
 except that instead of  $\xi( \mu^2, -E)=  \pi   \log (-E/\mu^2)$
they have the  function   (p.99, equation (5.16))  
\begin{equation}
   4 \pi^2  \alpha - 2 \pi \Psi(1)  + \pi \log (-E/4) 
\end{equation}
Comparing 
we see that they agree exactly if the parameters are related by 
\begin{equation}
\log \mu   =   - 2 \pi \alpha  +   \Psi(1)  + \log 2
 \end{equation}
\end{enumerate}

\subsection{extension}

The previous results can be generalized to allow an additional 
potential besides the delta function.   We  consider
\begin{equation}  
  H^\#  = -\De + v - g \de 
\end{equation}
For simplicity we will assume $v$ is a bounded function on  $\bbR^2$.
To define this we again start with approximate Hamiltonians in momentum space
\begin{equation} 
 H^\#_{\La}  = H_0  +v'  -  (2\pi)^{-2} g_{\La}\ P_{\rho_{\La}} 
 \end{equation}
where     $g_{\La} = g_{\La} (\mu)$  is  as before   and  $v'= \cF v\cF^{-1}$ is
a convolution operator  ($\cF$ = Fourier transform).  Since  $\|v'\| = \|v\| = \|v\|_{\infty}$
this is
still a bounded  perturbation and so $ H^\#_{\La}$ is  self-adjoint on  $D(H_0)$.
Without the approximate delta function we have
\begin{equation}
H_1 = H_0  +v'  
\end{equation}
which is also self-adjoint on  $D(H_0)$ and satisfies
and   $H_1 \geq  -\|v\|_{\infty}$.

Resolvents are denoted  
\begin{equation}
R_1(E)  = (H_1-E)^{-1}\ \ \ \ \ \ \  R^\#_{\La}(E)  =  (H^\#_{\La}-E)^{-1}
\end{equation}
If $E$ is complex and not in    $[ - \|v\|_{\infty}, \infty)$ then $E$
is in the resolvent set for  $H_1$.   As before we find that    
that  such   $E$ are also  in the resolvent set for $ H^\#_{\La}$ if and only 
if   $(2\pi)^2 g_{\La}^{-1}  \neq   (\rho_{\La},
R_1(E)  \rho_{\La})$  in which case 
\begin{equation}  \label{rr}
 R^\#_{\La}(E) 
 =  R_1(E) 
 + \left(\frac {1}{   (2\pi)^2 g_{\La}^{-1}-  (\rho_{\La},
R_1(E)  \rho_{\La})}
\right) P_{R_1(E)  \rho_{\La}} 
\end{equation}

\begin{thm} \ \   \label{one-a}
\begin{enumerate}
\item 
 For real  $E  <  - e_0$  with  
  \begin{equation}
 e_0  =  \max ( \|v\|_{\infty} +1,  \mu^2 e^{\|v\|_{\infty} +1}  )
\end{equation}
the strong limit   $R^\#(E) = \lim_{\La \to \infty} R^\#_{\La}(E)$ exists.
\item   $R^\#(E)$ is invertible.
\item   $R^\#(E) = \lim_{\La \to \infty} R^\#_{\La}(E)$ exists
for all complex   $E$ not in  $[-e_0, \infty) $.  
 There is a self-adjoint operator
$H^\#(\mu) $  satisfying $H^\#(\mu)  \geq -e_0 $  such that  
 \begin{equation}
 R^\#(E)  =  (H^\#(\mu) -E)^{-1}
\end{equation}
\end{enumerate}
\end{thm}
\bigskip

\pr
In the   denominator in (\ref{rr}) we insert
\begin{equation}
 R_1(E)=R_0(E)-R_1(E) v' R_0(E)
\end{equation}
and find  
\begin{equation}
 R^\#_{\La}(E)= R_1(E) 
 + \left(\frac {1}{   \xi_{\La}(\mu^2,-E) +  (\rho_{\La},R_1(E) v' R_0(E)\rho_{\La})}
\right) P_{R_1(E) \rho_{\La}} 
\end{equation}
As  $\La  \to \infty$  we have  in  $\cL^2(\bbR^2)$ 
\begin{equation}
\begin{split}
\lim_{\La \to \infty} R_1(E) \rho_{\La} 
= & \lim_{\La \to \infty} R_0(E) \rho_{\La} -R_1(E) v' R_0(E) \rho_{\La} \\
= &  \Om_E   -R_1(E) v' \Om_E\\ 
 \equiv & \Om_{1,E}\\
\end{split}
\end{equation}
Thus we have  the limit  $ R^\#(E)=  \lim_{\La \to \infty}  R^\#_{\La}(E)$
given by  
\begin{equation}
 R^\#(E)= R_1(E) 
 + \left(\frac {1}{   \xi(\mu^2,-E) +  (\Om_E, v' \Om_E)  - (\Om_E  ,v'R_1(E) v' 
\Om_E)        }
\right) P_{\Om_{1,E}} 
\end{equation}
provided the denominator does not vanish.
However     $ \|\Om_E\|_2^2  =  \pi |E|^{-1}$ and     since  $ E < -  \|v\|_{\infty} -1$ we
have 
 $\|R_1(E)\|\leq 1$ and hence 
\begin{equation}
\begin{split}
 |(\Om_E, v' \Om_E)| \leq &   \pi|E|^{-1} \|v\|_{\infty}  \leq   \pi\\
| (\Om_E  ,v'R_1(E) v'  \Om_E) | \leq &  \pi|E|^{-1} \|v\|^2_{\infty}  \leq   \pi \|v\|_{\infty} \\
\end{split}
\end{equation}
Thus we can avoid vanishing provided   
$ \xi(\mu^2,-E)  >   \pi (\|v\|_{\infty}+1)   $   or  $\log(-E/\mu^2) >\|v\|_{\infty}+1 $.  This is
our condition
 $ -E > \mu^2  e^{\|v\|_{\infty} +1}  $.  

Thus part one is proved.   The second and third parts 
 follow as in the previous theorem.

\section{Many particles}

\subsection{bosons}
We now turn to the many particle problem.   It is convenient to work
with all possible values of $N$ at the same time,  even though
the main interest is at fixed   $N$.  This means we are working 
on the Fock space $\cH =   \oplus_{N=0}^{\infty} \cH_N$.
This has the usual creation and annihilation operators $a^*(f), a(f)$
defined for  $f \in \cL^2(\bbR^2)$.  We also have  $a(p) = a(\de(\cdot-p))$
defined on the domain  $\cD$ which is the dense subspace 
of $\cH$ with  only  a finite number of entries and
wave functions in the Schwartz space  $\cS(\bbR^2)$.
For  $\psi \in \cD$ the function  $p \to  a(p) \psi$  is rapidly decreasing.
(Note that   $a^*(p) = a^*(\de(\cdot-p))$
is not an operator.)

The  Hamiltonian has the form    
$H = H_0   + H_{I}$.   
The free Hamiltonian $ H_0 $ is    $\sum_{i=1}^N p_i^2/2 $ on the 
$\cH_N$  and  is essentially self-adjoint on  $\cD \cap \cH_N$. It can also  be represented as a
bilinear  form  on  $\cD \times \cD$ as
\footnote{ This means 
 $ (\phi,H_0 \psi)  =  \int  \om(p) (a(p) \phi, a(p) \psi) dp  $
or as a quadratic form  
 $ (\psi,H_0\psi)  =  \int   \om(p)  \| a(p)\psi\|^2 dp  $}
\begin{equation}
H_0  =   \int   \om(p)  a^*(p) a(p)  dp \ \ \ \ \ \ \ \   \om(p) = \frac{p^2}{2}
\end{equation}
The interaction  with  interparticle potential  $-g\de(x-y)$   is given in momentum space
by  the  bilinear form on   $\cD\times \cD$:
\begin{equation}
H_{I} = \frac{-g}{2 (2 \pi)^2}  \int
 a^*(p_1') a^*(p_2')\delta( p_1 + p_2 -p_1'-p_2')
a(p_1) a(p_2)\ dp_1 dp_2  dp'_1 dp'_2   
\end{equation}
However this is not an operator.

 To remedy this we 
introduce
\begin{equation}
H_{\La}   =  H_0 +  H_{I,\La}
\end{equation}
For   $ H_{I,\La}$ we   
add momentum
cutoffs $\rho_{\La}$, take the coupling constant  $g_{\La} = g_{\La}(\mu)$
as before, and define 
\begin{equation}
\begin{split}
H_{I,\La} =& \frac{-g_{\La}}{2 (2 \pi)^2}  \int 
\rho_{\La}(\frac{p_1 - p_2}{2}) \rho_{\La}(\frac{p_1'-p_2'}{2})  \\
& a^*(p_1') a^*(p_2')\delta( p_1 + p_2 -p_1'-p_2')
a(p_1) a(p_2)\ dp_1 dp_2  dp'_1 dp'_2  \\
\end{split}
\end{equation}
Changing variables to  
\begin{equation}
p=p_1+p_2   \ \ \ \ \ \ \ \ q=\frac{p_1-p_2}{2}
\end{equation}
we find for the associated quadratic form
\begin{equation}  \label{rep}
\begin{split}
(\psi,H_{I,\La}\psi ) 
 =&  \frac{-g_{\La}}{2 (2 \pi)^2}    \int
\rho_{\La}(q) \rho_{\La}(q') \\
&
 \left(a(\frac{p}{2}+q') a(\frac{p}{2}-q')\psi,
a(\frac{p}{2}+q) a(\frac{p}{2}-q )\psi\right)\ dpdqdq'  \\
\end{split}
\end{equation}
 Applying
the Schwarz inequality first in Fock space and then in the integral
we find 
\begin{equation}
\begin{split}
|(\psi,H_{I,\La}\psi )| 
 \leq   &  \frac{g_{\La}}{2 (2 \pi)^2}  
\left( \int  \rho_{\La}(q)^2 dq  \right)
\int  \|a(\frac{p}{2}+q) a(\frac{p}{2}-q )\psi\|^2  dp dq  \\
=    &  \frac{g_{\La}}{2 (2 \pi)^2}  
  \|\rho_{\La}\|_2^2
\int  \|a(p_1) a(p_2 )\psi\|^2  dp_1 dp_2  \\
=     &  \frac{g_{\La}}{2 (2 \pi)^2}  
  \|\rho_{\La}\|_2^2
 \|N_0^{1/2}(N_0-1)^{1/2}\psi\|^2   \\
\end{split}
\end{equation}
Here  $N_0 =   \int a^*(p) a(p)  dp$ is the number operator.

On the $N$-particle subspace  $\cH_N$
we have   $N_0=N$  and hence $H_{I,\La}$ is a bounded quadratic form.   This
determines a bounded self-adjoint operator on each  $\cH_N$ and hence     $H_{\La}$ defines a  
self-adjoint operator  on each $\cH_N$ with  domain  $D(H_0)   \cap   \cH_N$.  Taking the
direct  sum we get a self-adjoint operator    $H_{\La}$ on the full Fock space.

\subsection{angels}
Next we introduce angels.  
We define 
\begin{equation}
\tilde \cH  =  \cL^2(\bbR^2)  \otimes  \cH 
\end{equation}
which is Fock space with an angel.   
For $f  \in \cL^2(\bbR^2)$ we define 
 $\chi(f): \tilde  \cH  \to  \cH $  and $\chi^*(f):  \cH  \to  \tilde  \cH$
by  
\begin{equation}
\begin{split}
\chi(f) (h \otimes \psi)   =&  (f,h)\psi \\
\chi^*(f) \psi =& f \otimes  \psi \\
\end{split}
\end{equation}
These are   creation and annihilation operators for angels, they are adjoint to 
each other,  and they
satisfy
\begin{equation}
\begin{split}
\chi(f) \chi^*(h) =& (f,h) \\
\chi^*(h) \chi(f)  =& h(f, \cdot)  \otimes I\\
\end{split}
\end{equation}
There is also the operator   $\chi(p) =  \chi( \de ( \cdot - p))$
defined say on the dense  subspace  $ \tilde \cD   \subset \tilde \cH$
defined by $
\tilde \cD  \equiv \cS(\bbR^2)  \otimes  \cD$.

An equivalent  representation is   
\begin{equation}  \label{alt}
\tilde \cH  =   \cL^2(\bbR^2,  \cH) 
\end{equation}
Then    $\tilde \cD$ is a subspace of  $\cS(\bbR^2,  \cD)$ and 
on this domain
\begin{equation}
\chi(p)  \Psi  =  \Psi(p)
\end{equation}

Next we introduce:
\begin{defn}  
\begin{equation}
B_{\La}  =  \frac{1}{\sqrt{2} (2 \pi)}    \int   \rho_{\La}(\frac{p_1-p_2}{2})  \chi^*(p_1+p_2) 
a(p_1) a(p_2)  dp_1 dp_2  
\end{equation}
\end{defn}

Then  $B_{\La}$ is an operator from  $\cH$ to  $\tilde \cH$, and 
the  key point is that  it  provides a square root for  $H_{I, \La}$.

\begin{lem}
  For  $\La < \infty$  
\begin{enumerate}
\item  $B_{\La}$ defines  a bounded operator on each subspace   $\cH_N$.
\item  For   $\psi \in \cD$ we have in the representation (\ref{alt})
\begin{equation}  \label{another}
(B_{\La}\psi)(p)  = \frac{1}{\sqrt{2} (2 \pi)}    \int \rho_{\La}(q)\ 
 a(\frac{p}{2}+q) a(\frac{p}{2}-q )\psi   \ dq
\end{equation}
\item On each  $\cH_N$:
 \begin{equation}  \label{sq}
- g_{\La}B^*_{\La} B_{\La}  =  H_{I, \La}
\end{equation}
\end{enumerate}
\end{lem}
\bigskip

\pr  The expression is naturally defined as a bilinear form.  For  $\psi \in \cD$
and  $\Psi \in  \tilde \cD $  we have in the representation (\ref{alt})
\begin{equation}  \label{b}
(\Psi, B_{\La} \psi)  =  \frac{1}{\sqrt{2} (2 \pi)}     \int \rho_{\La}(\frac{p_1-p_2}{2}) (
\Psi(p_1+p_2),  a(p_1) a(p_2)\psi)  dp_1 dp_2 
\end{equation}
Applying the Schwarz inequality twice 
we have 
\begin{equation}   \label{sch}
\begin{split}
|(\Psi, B_{\La} \psi)|  \leq  &  \left(  \int |\rho_{\La}(\frac{p_1-p_2}{2})|^2
 \| \Psi(p_1+p_2)\|^2  dp_1 dp_2  \right)^{1/2}
\left (\int \| a(p_1) a(p_2)\psi \|^2  dp_1 dp_2 \right ) ^{1/2}\\
\leq &  \|\rho_{\La}\|_2
 \| \Psi\|  \| N_0 \psi\| 
\end{split}
\end{equation}
Now specialize to $\psi \in \cH_N$  and we see 
that $B_{\La} $ is a bounded bilinear form and hence a bounded operator.
This establishes  the first point.

Next change variables  in  (\ref{b}) and obtain
\begin{equation}  
(\Psi, B_{\La} \psi)  =   \frac{1}{\sqrt{2} (2 \pi)}    \int \rho_{\La}(q) \left( \Psi(p), 
  a(\frac{p}{2}+q) a(\frac{p}{2}-q)\psi \right) dp dq 
\end{equation}
which establishes  (\ref{another}).

For  (\ref{sq}) it suffices to establish the identity as a quadratic form
  on   $\cD$.
 Inserting the representation  of (\ref{another})  into  $- g_{\La}\| B_{\La}\psi\|^2$
we obtain the representation (\ref{rep})  of   $(\psi, H_{I,\La} \psi)$. 
This completes the proof.
\bigskip

For later reference we consider the case   $\La = \infty$ with the operator 
\begin{equation}
B  =  \frac{1}{\sqrt{2} (2 \pi)}    \int  \chi^*(p_1+p_2)  a(p_1) a(p_2)  dp_1 dp_2  
\end{equation}

\begin{lem}  \label{B}
$B$ defines an (unbounded)  operator on  $\cH_N  \cap D(H_0)$
which satisfies for some constant $C$:
\begin{equation}   \label{un}
 \|   B \psi\| \leq  C \|(H_0+N ) \psi\| 
\end{equation}
For   $\psi$ in this domain
\begin{equation}
\lim_{\La \to \infty}  B_{\La} \psi  =  B \psi
\end{equation}
\end{lem}
\bigskip

\pr
All the above representations still hold for   $\psi \in \cD,  \Psi \in \tilde \cD$.  But now instead
of  (\ref{sch}) we have:
\begin{equation}  \label{sch2}
\begin{split}
|(\Psi, B \psi)|  \leq  &  \left(  \int  (\om(p_1) +1)^{-1} (\om(p_2) +1)^{-1}
 \| \Psi(p_1+p_2)\|^2  dp_1 dp_2  \right)^{1/2}\\
 \times  &
\left (\int   (\om(p_1) +1) (\om(p_2) +1) \| a(p_1) a(p_2)\psi \|^2  dp_1 dp_2 \right ) ^{1/2}\\
\leq  &  \left(  \int  (\om(\frac{p}{2} + q +1)^{-1} (\om(\frac{p}{2} +q) +1)^{-1}
 \| \Psi(p)\|^2  dp dq  \right)^{1/2} \|(H_0 + N_0) \psi\| \\
 \leq &  C
 \| \Psi\|  \|(H_0 + N_0) \psi\| 
\end{split}
\end{equation}
Here  $C =  (\int (\om(q)+1)^{-2} dq)^{1/2}$ and   in the last step we use the Schwarz inequality
in 
$q$. This shows that   $B$ defines an operator on  $\cD \cap \cH_N$ satisfying the 
inequality (\ref{un}).   Since $\cD \cap \cH_N$ is a core for  $H_0$ on $\cH_N$ we can extend the domain
to  $D(H_0) \cap \cH_N$.  

For the second point we estimate $|(\Psi, (B-B_{\La}) \psi)|$ as above.  In the last integral
over $q$ we are  now restricting to  $|q| \geq \La $.  Break this into two terms using
\begin{equation}
\{ q:  |q| \geq \La \}  \subset \left \{ q:  |\frac{p}{2}+q| \geq  \frac{\La}{2} \right
\}  
\cup  \left \{ q:  |\frac{p}{2}-q| \geq  \frac{\La}{2} \right\}  
\end{equation}
With    $\de C_{\La} =  (\int_{q \geq \La} (\om(q)+1)^{-2} dq)^{1/2}$  and
$\psi  \in  \cD \cap \cH_N$ this leads   
\begin{equation}  \label{ccc}
|(\Psi, (B_{\La} -B) \psi)|  \leq \sqrt{ 2C_{\La} C}  
 \| \Psi\|  \|(H_0 + N) \psi\| 
\end{equation}
This estimate extends  to  $\psi   \in  D(H_0) \cap \cH_N$.
Then as $\La \to 0$
we have   $\de C_{\La} \to 0$ and  $ \|(B_{\La} -B) \psi\| \to 0$.

\subsection{resolvents}

We return to  $\La < \infty$ and  work out some consequences of the  identity (\ref{sq})
   for    resolvents.    We define  
\begin{equation}
R_0(E)  = (H_0-E)^{-1}\ \ \ \ \ \ \  R_{\La}(E)  = (H_{\La}-E)^{-1}
\end{equation}
These exist for $\textrm{Im}E \neq 0$ and 
   $R_0(E)$  exists for $E   <0$.
We  want to find real E such that  $R_{\La}(E)$   exists as a means to isolate 
the spectrum of $H_{\La}$.

To this end we  also introduce the  operators on  $\cH \oplus \tilde  \cH$
\begin{equation}
\tilde H_{\La}(E)  = \left(   \begin{array}{cc}
H_0 -E   &   B^*_{\La}  \\
 B_{\La} &  g_{\La}^{-1} \\
\end{array}   \right) \ \ \ \ \ \   \tilde R_{\La}(E)  =  \tilde H_{\La}(E)^{-1} 
\end{equation}
Since  $B_{\La}$ is a bounded operator from   $\cH_N$ to  $\tilde \cH_{N-2}$
we have that   $\tilde H_{\La}(E)$ preserves the subspace   
 $\cH_N \oplus \tilde  \cH_{N-2}$. More precisely it is defined on   
$( D(H_0)  \cap \cH_N) \oplus \tilde  \cH_{N-2}$ and is self-adjoint there.

\begin{lem}
For $E<0$, $R_{\La}(E)$ exists  in $\cB(\cH_N)$ iff $\tilde R_{\La}(E)$ exists  in  
$\cB(\cH_N\oplus \tilde \cH_{N-2})$  
 in which case 
 \begin{equation}   \label{inverse}
\tilde R_{\La}(E)   = \left(   \begin{array}{ccc}
R_{\La}(E)   &   & -g_{\La}R_{\La}(E) B^*_{\La} \\
 -g_{\La}B_{\La}R_{\La}(E) &  &  g_{\La}  +  g_{\La}^2   B_{\La}R_{\La}(E)  B^*_{\La}   \\
\end{array}   \right)
\end{equation}
\end{lem}
\bigskip

\pr  We omit the subscript $\La$ for the proof. First assume that $\tilde R(E)$ exists. 
 Then it is self-adjoint and has the form 
\begin{equation}
\tilde R(E)   = \left(   \begin{array}{ccc}
\al  &   &\beta^*  \\
\beta &  &  \delta \\
\end{array}   \right)
\end{equation}
for bounded  $\al, \beta, \delta $ and  $\al, \de $  self-adjoint.  The statement that
is   the inverse says  that  $\al, \beta^*$ map into the domain of  $H_0$ and that
\begin{equation}  \label{array}
\begin{split}
(H_0 - E)\al  + B^* \beta =& I  \\
(H_0 - E)\beta^*  + B^* \de =& 0  \\
B\al  + g^{-1} \beta =& O  \\
B \beta^*  + g^{-1}  \de =& I  \\
\end{split}
\end{equation}
We ignore the second equation.  The third equation says 
\begin{equation}
 \beta =  -g B\al  \ \ \ \ \ \   \beta^* =  -g \al B^*
\end{equation}
Inserting the expression for $\beta $ into the first equation 
and using  $- gB^* B = H_I $  we get $(H-E) \al = I$.
Hence $  R(E)$ exists  and equals  $\al$. Inserting the expression for $\beta^* $ into the last
equation  gives  $\de  =  g + g^2 g B  R(E) B^*$

On the other hand if  $R(E)$ exists one can check directly that (\ref{inverse}) provides 
a bounded inverse.  This completes the proof.
\bigskip

Now we give another version.

\begin{defn}   For  $E<0$  define a bounded operator on  each $\tilde \cH_N$ by 
\begin{equation}
\Phi_{\La} (E)  =  g_{\La}^{-1}  - B_{\La}R_0(E) B_{\La}^*
\end{equation}
\end{defn}

\begin{lem}
For $E<0$,   $\tilde R_{\La}(E)$ exists  in  $\cB( \cH_N \oplus\tilde \cH_{N-2})$ iff 
$\Phi_{\La} (E) ^{-1}$ exists  in $\cB( \tilde  \cH_{N-2})$ in which case 
 \begin{equation}   \label{inverse2}
\tilde R_{\La}(E)   = \left(   \begin{array}{ccc}
R_0(E)  + R_0(E) B^*_{\La}\Phi_{\La} (E) ^{-1} B_{\La}R_0(E)   &   & -R_0(E)B^*_{\La}
\Phi_{\La}(E)^{-1}
\\
-\Phi_{\La}(E)^{-1}B_{\La}R_0(E)
&  & \Phi_{\La} (E) ^{-1} \\
\end{array}   \right)
\end{equation}
\end{lem}
\bigskip

\pr   Again suppose that $\tilde R(E)  $ exists so we must solve the equations
(\ref{array}) again. This time we ignore the third equation.  Then the second 
equation says that  
\begin{equation}
\beta^*  = - R_0(E) B^* \de  \ \ \ \ \ \ \ \ \beta  = - \de B R_0(E)
\end{equation}
Substituting $\beta^*$ into the fourth equation  gives
$(-BR_0(E) B^*   + g^{-1})  \de = I  $  or   $\Phi(E) \de  =I$.  Hence    
$\Phi(E)^{-1}$ exists   and equals $\de$.  Substituting  $\beta$  into  the first
equation gives  $(H_0 - E)\al  - B^* \Phi(E)^{-1}  B R_0(E) = I $ whence     
$ \al  =  R_0(E)  + R_0(E)   B^* \Phi(E)^{-1}  B R_0(E) $.

On the other hand if  $\Phi(E)^{-1}$ exists one can check directly that (\ref{inverse2}) provides 
a bounded inverse.  This completes the proof.
\bigskip

Comparing these results we have: 
\begin{lem}   \label{equiv}
For $E<0$, $ R_{\La}(E)$ exists  in  $\cB( \cH_N)$ iff 
$\Phi_{\La} (E) ^{-1}$ exists  in $\cB(\tilde \cH_{N-2})$ in which case 
 \begin{equation}  
\begin{split}
 R_{\La}(E)   = & R_0(E) + R_0(E) B^*_{\La}\Phi_{\La} (E) ^{-1} B_{\La}R_0(E) \\
 \Phi_{\La} (E) ^{-1} =&  g_{\La}  +  g_{\La}^2   B_{\La}R_{\La}(E)  B^*_{\La}  \\
\end{split}
\end{equation}
\end{lem}
\bigskip

\subsection{renormalization}

In view of the last result we can study the resolvent  $R_{\La}(E)$  on $\cH_N$ by 
studying the operator $\Phi_{\La}(E)$ on  $\tilde \cH_{N-2}$.  The advantage
of this operator is that it can be more  easily renormalized.

First we Wick order moving creation operators to the left and annihilation
operators to the right using  $[a(p),a^*(p')] = \de(p-p')$ and 
\begin{equation}
(H_0 -E)^{-1} a^*(p) =  a^*(p) (H_0 + \om(p)  -E)^{-1}
\end{equation}
The resulting identity is formal but a rigorous version can be had by
regularizing  $a(p) \to a(\de_{\ka}(\cdot - p))$ with approximate
delta functions  $\de_{\ka}$. 
We find 
\begin{equation}
\Phi_{\La}(E) = \Phi_{0,\La}(E) + \Phi_{I,\La}(E)
\end{equation}
where  
\begin{equation}
\begin{split}
\Phi_{0,\La}(E)= & g_{\La}^{-1}  -  \frac{1}{2 (2 \pi)^2}    \int dp_1 dp_2 \ \chi^*(p_1 +
p_2)
\chi(p_1+p_2)
\rho_{\La}(\frac{p_1-p_2}{2})^2   
 \frac {2}{H_0   + \om_1  +\om_2 -E } \\
\Phi_{I,\La}(E)= &  -  \frac{1}{2 (2 \pi)^2}   \int dp_1 dp_2  dp'_1 dp'_2 
\chi^*(p_1 + p_2) \chi(p_1'+p_2') \rho_{\La}(\frac{p_1-p_2}{2}) \rho_{\La}(\frac{p_1'-p'_2}{2}) \\
&\left(  a^*(p_1') a^*(p_2')\frac {1}{H_0 + \om_1  +\om_2  + \om'_1  +\om'_2-E} 
a(p_1) a(p_2) \right. \\
&\left. +\de (p_1-p_1')  a^*(p_2') \frac {4}{H_0  +\om_1  + \om_2  +\om'_2-E }  a(p_2)
\right)
\\
\end{split}
\end{equation}
Here  $\om_1=\om(p_1)=p_1^2/2$, etc.
These are bilinear forms on  $\tilde \cD  \times \tilde \cD$.
By the methods of section 3.2 they  determine bounded operators  on each  $\tilde \cH_N$
for  $\La < \infty$.  But now we want to work uniformly in  $\La$ and also 
include  $\La = \infty$.

To  cancel the  divergence in  $\Phi_{0,\La}(E)$
we change variables and write
\begin{equation}
\Phi_{0,\La}(E) =
 (2\pi)^{-2} \left( \int_{ |q| \leq  \La}  (q^2+ \mu^2)^{-1}
-
   \int_{ |q| \leq  \La} dpdq \ \chi^*(p) \chi(p) 
 \frac {1}{H_0   + p^2/4+q^2-E }\right) 
\end{equation}
In the representation  $\tilde \cH   =  \cL^2(\bbR^2, \tilde \cH )$
this is 
\footnote{ In general if $T =  \int \chi^*(p)T(p) \chi(p)  dp$  defines  an 
operator on  $\tilde \cH  = \cL^2(\bbR^2) \otimes  \cH $,  then in 
the representation   
$\tilde\cH
 =\cL^2(\bbR^2, \cH)$ we have   $(T \Psi)(p)  =  T(p) \Psi(p)$.  }
\begin{equation}
\begin{split}
(\Phi_{0,\La}(E) \Psi)(p) = &
 (2\pi)^{-2} \left( \int_{ |q| \leq  \La}  (q^2+ \mu^2)^{-1}
-
   \int_{ |q| \leq  \La} dq \
 \frac {1}{H_0   + p^2/4+q^2-E }\right) \ \Psi(p)\\
=  & (2\pi)^{-2}\xi_{\La}(\mu^2, H_0   + p^2/4-E)  \ \Psi(p) \\
\end{split}
\end{equation}
As noted in (\ref{cancel}),   $\xi_{\La}$ has no divergence and we 
can define for  $\La = \infty$:
\begin{equation}
\begin{split}
(\Phi_{0}(E) \Psi)(p)=& (2\pi)^{-2}\xi(\mu^2, H_0   + p^2/4-E) \ \Psi(p)\\
=& (4\pi)^{-1}   \log\left(\frac{ H_0   + p^2/4-E}{\mu^2}  \right) 
\Psi(p) \\
\end{split}
\end{equation}

\begin{lem}  For  $E < - \mu^2 $,  
$\Phi_0(E)$ is essentially self-adjoint on
$\tilde \cD \cap \tilde  \cH_N$ and for  $\Psi$ in this domain
we have   
\begin{equation}   \label{zero}
\lim_{\La \to \infty } \Phi_{0,\La}(E) \Psi =   \Phi_{0}(E) \Psi  
\end{equation}
\end{lem}

\pr  
For the essential self-adjointness it suffices to show that the domain 
contains a dense set of analytic vectors. (Nelson's theorem,  \cite{RS75}).
For analytic vectors we can take wavefunctions with compact support.

The convergence is straightforward.  One can use the inequality 
\begin{equation}
\begin{split}
&\| \left( \log(H_0 + p^2/4 -E + \frac{1} {\La^2}) 
  -  \log(H_0 + p^2/4 -E )\right)\Psi(p)\|\\
\leq & \ 
\La^{-2}\|(H_0 + p^2/4 -E )^{-1}\Psi(p)\|
\leq \cO( \La^{-2})\|\Psi(p)\| \\
\end{split}
\end{equation}
which follows using the spectral theorem.
\bigskip

Next we work on  $\Phi_{I,\La}(E)$.  For $\La= \infty$ it is defined without the $\rho_{\La}$ and denoted
$\Phi_{I}(E)$.

\begin{lem}  \label{last} For   $E< -1$ and  $\La \leq \infty$  and  $\Psi \in \tilde \cD$:
\begin{equation}  \label{xy}
|(\Psi, \Phi_{I,\La}(E)  \Psi) | \leq 2  (\Psi, N^2_0 \Psi)
\end{equation}
Thus   $ \Phi_{I,\La}(E)  , \Phi_{I}(E)  $ define bounded operators on $ \tilde \cH_N$
and for  
$\Psi \in \tilde \cD \cap \tilde \cH_N$:
 \begin{equation}  \label{I}
\lim_{\La \to \infty } \Phi_{I,\La}(E) \Psi =   \Phi_{I}(E) \Psi  
\end{equation}
\end{lem}

\pr    We take   $ \Phi_{I,\La}(E)
=  \Phi^{(2)}_{I,\La}(E)  + \Phi^{(4)}_{I,\La}(E) $ where the superscript indicates 
the number of creation and annihilation operators. 
For the first we have   
\begin{equation}
\begin{split}
&|(\Psi, \Phi^{(2)}_{I,\La}(E)  \Psi)| \\ 
 \leq   & \frac{1}{2\pi^2}
 \int dp_1 dp_2  dp'_2 
| \left( a(p_2')  \Psi(p_1+p_2),\frac {1}{H_0   + \om_1  +\om_2   +\om'_2-E}  a(p_2)  
\Psi(p_1+p_2') \right)|\\
 \leq   &  \frac{1}{2\pi^2}
 \int dp_1 dp_2  dp'_2 
 \| a(p_2')  \Psi(p_1+p_2)\|\frac {1}{ \om_2   +\om'_2 +1} 
\| a(p_2)  \Psi(p_1+p_2') \| \\
 \leq   &  \frac{1}{2\pi^2}
 \int dp_2  dp'_2 
 \| a(p_2')  \Psi\|\frac {1}{ \om_2   +\om'_2 +1} 
\| a(p_2)  \Psi \| \\
\leq &  \| N_0^{1/2}  \Psi \|^2 
\end{split}
\end{equation}
Here in the last step we use the fact,  noted in  \cite{DFT94},
that  for  $h,h' \in \cL^2(\bbR^2)$ and any $c>0$:
\begin{equation}  \label{x}
|\int h(p)\frac{ 1}{p^2 +q^2 +c}  h'(q) dp dq|   \leq \pi^2 \|h\|_2\|h'\|_2 
\end{equation}

For the convergence we proceed differently.  We use the 
estimate for  $ \ep >0$
\begin{equation}
 |\rho_{\La}(\frac{p_1-p_2}{2})\rho_{\La}(\frac{p'_1-p'_2}{2})-1|\leq \cO(\La^{-\ep})
(\om_1 +\om_2  +\om'_2 +1)^{\ep} 
\end{equation}
Then  
for  $\Psi_1,\Psi_2 \in \tilde \cD$ we have   
\begin{equation}  \label{convergence}
\begin{split}
&|(\Psi_1, (\Phi^{(2)}_{I,\La}(E) -  \Phi^{(2)}_{I}(E)) \Psi_2)| \\ 
 \leq   &  \cO(\La^{-\ep})
 \int dp_1 dp_2  dp'_2 
 \| a(p_2')  \Psi_1(p_1+p_2)\|\frac {1}{(\om_1 +\om_2  +\om'_2 +1)^{1-\ep}} 
\| a(p_2)  \Psi_2(p_1+p_2') \| \\
 \leq   &  \cO(\La^{-\ep})
 \int dp_1 dp_2  dp'_2 \frac {1}{(\om_2  +1)^{\frac34-\ep/2}}
 \| a(p_2')  \Psi_1(p_1+p_2)\| \frac {(\om_2  +1)^{1/2}}
{(\om'_2 +1)^{\frac34-\ep/2}} 
\| a(p_2)  \Psi_2(p_1+p_2') \| \\
\leq &  \cO(\La^{-\ep}) \|N_0^{\frac12} \Psi_1\|\| (H_0 +N_0)^{\frac12}  \Psi_2 \||
\end{split}
\end{equation}
where the last step follows by the Schwarz inequality.  
Specializing to  $\tilde \cD \cap \tilde  \cH_N$ the estimate is uniform in  $\|\Psi_1\| = 1$ 
and yields the convergence    
$\|(\Phi^{(2)}_{I,\La}(E) -  \Phi^{(2)}_{I}(E)) \Psi_2\| \to  0 $
(In fact  strong convergence holds since we have a uniform bound on the norms).

For the second term we define
\begin{equation}
 f(p',q',p)  = \|a(\frac{p'}{2}+q') a(\frac {p'}{2}-q') \Psi(p)\|
\end{equation}
and find 
\begin{equation}
\begin{split}
&(\Psi, \Phi^{(4)}_{I,\La}(E)  \Psi)   \leq   \frac{1}{8\pi^2}  \int dp_1 dp_2  dp'_1 dp'_2 
\\
&\|a(p_1') a(p_2') \Psi(p_1+p_2)\|\frac {1}{ \om_1  +\om_2  + \om'_1 
+\om'_2+1} \| a(p_1) a(p_2)\Psi(p_1'+p_2') \| \\
\leq  &  \frac{1}{8\pi^2}  \int dp dq dp'
dq'\  f(p',q',p) \frac {1}{ q^2  + (q')^2+1}   f(p,q,p') \\ 
\leq & \frac{1}{8} \int dp dp' \| f(p',\cdot ,p) \|_2  \| f(p,\cdot ,p') \|_2   \\
\leq & \frac{1}{8}  \|f\|_2^2  
\leq  \frac{1}{8}   \| N_0 \Psi \|^2      \\
\end{split}
\end{equation}
Again we have used  (\ref{x}).
This completes the bound, and the convergence follows by an estimate similar to   (\ref{convergence})  
\bigskip

To combine these we have   :

\begin{lem}  { \ }  \label{verylast}
\begin{enumerate}
\item  For  $E  < - 1$,  
$\Phi(E)$  is essentially self-adjoint
on  $\tilde \cD  \cap  \tilde  \cH_N$  and 
 for     $\Psi$  in this domain
\begin{equation}
\lim_{\La \to \infty}  \Phi_{\La}(E) \Psi =   \Phi(E) \Psi
\end{equation}
\item  Let  $E  <  -e_N$
where   
\begin{equation}
e_N =  \max(1,   \mu^2 e^{ 16 \pi N^2})
\end{equation}
    Then for  $\La$  sufficiently large  or  $\La = \infty$ we have that     $\Phi_{\La}(E)$
is strictly positive and for 
  $\Psi\in \tilde \cH_N$
\begin{equation}
\lim_{\La \to \infty}  \Phi_{\La}(E)^{-1} \Psi =   \Phi(E)^{-1} \Psi
\end{equation}
\end{enumerate}
\end{lem}

\pr  $\Phi (E)  = \Phi_{0} (E) + \Phi_{I} (E)$
is the sum of a essentially    self adjoint operator and a bounded 
operator.   The  essential  self-adjointness again  follows by  
 Kato's  theorem.  
The convergence follows from our results   (\ref{zero}), (\ref{I}).

For the second part  under our assumptions
 $\xi(\mu^2,-E) =\pi  \log (-E/\mu^2)   \geq   16 \pi^2 N^2$. 
Then  since  
  $\xi_{\La}(\mu^2,-E)$ converges to   
 $\xi(\mu^2,-E) $
 we have  for $\La$ sufficiently large    (depending on  $E, \mu$)  
 $\xi_{\La}(\mu^2,-E)   \geq   12 \pi^2 N^2$.
Since $\xi_{\La}(a,b)$ is increasing in $b$ we have  for  $\Psi \in \tilde \cD  \cap  \tilde  \cH_N$: 
\begin{equation}  \label{ee}
 (\Psi, \Phi_{0,\La}(E) \Psi) \geq   (2\pi)^{-2}  \xi_{\La}(\mu^2,-E)\| \Psi\|^2
\geq    3N^2 \| \Psi\|^2 
\end{equation}
  Combining this with the bound $ |(\Psi, \Phi_{I,\La}(E) \Psi)| \leq  2N^2\| \Psi\|^2$ we have
for  $\La$ sufficiently large or  $\La = \infty$:
\begin{equation}
 (\Psi, \Phi_{\La}(E) \Psi)
\geq    N^2 \| \Psi\|^2 
\end{equation}
This gives the positivity and shows that  
  $\Phi_{\La}(E)$ has a bounded inverse.    
 Convergence on   the  core   $ \tilde \cD   \cap \tilde \cH_N $  for  $\Phi(E)$  and the  uniform bound  $\| 
\Phi_{\La}(E) ^{-1}
\|
\leq  N^{-2}$ imply the strong convergence for   $\Phi_{\La}(E)^{-1}$. (See for example \cite{Kat66},  p.429) 
\newpage

\subsection{resolvent convergence}

Now we can prove the main result  (c.f. Dell'Antonio, Figari,Teta \cite{DFT94} )

\begin{thm} { \ }  \label{three}
\begin{enumerate}
\item 
For real   $E < - e_N$ and  $\psi \in \cH_N$
the limit  $R(E)\psi   = \lim_{\La \to \infty} R_{\La}(E) \psi $ 
exists and is equal to  
\begin{equation}  \label{resolvent2}
R(E) 
 =  R_0(E)  + R_0(E) B^*\Phi (E) ^{-1} B  R_0(E)
\end{equation}
\item   $R(E)$ is invertible.
\item  For  $E$ complex and not in $[-e_N, \infty)$ the limit  
 $R(E)\psi   = \lim_{\La \to \infty} R_{\La}(E) \psi $ exists.
There  is a self-adjoint operator  $ H(\mu)$  with  
 $H(\mu)  \geq   -e_N$ so 
$R(E) = (H(\mu)-E)^{-1}$.
\end{enumerate}
\end{thm}

\pr 
\begin{enumerate}
\item  By lemma \ref{verylast} if  $E < -e_N$ and $\La
$ is sufficiently large
 then    $\Phi_{\La}(E)^{-1} $     
exists as a bounded operator on $\tilde \cH_{N-2}$.  By lemma \ref{equiv} it follows
that all such real $E$ are in the resolvent set of  $H_{\La}$ on $\cH_N$
and 
\begin{equation}
 R_{\La}(E)   =  R_0(E)  + R_0(E) B^*_{\La}\Phi_{\La} (E) ^{-1} B_{\La}R_0(E)
\end{equation}

We claim that  $ B_{\La}R_0(E)$  converges in norm  to  $ BR_0(E)$.
By the resolvent identity it suffices to prove this for any  $E<0$ and 
we take  $E = -N$ and show   $ B_{\La}(H_0 +N)^{-1}  $  converges in norm  to  $ B(H_0 +N)^{-1} $.   
This follows by  (\ref{ccc}). 
Taking adjoints we have   that   $R_0(E)B^*_{\La}$     converges in norm 
to    $R_0(E)B^*$.  We also know by  lemma \ref{verylast} that 
  $\Phi_{\La} (E) ^{-1} $ converges 
strongly to  $\Phi (E) ^{-1}$.  Combining these results  we have that 
 $ R_{\La}(E) $   converges strongly to   $R(E)$ given by  (\ref{resolvent2}).

\item   
As in the proof of theorem \ref{one} it  suffices to 
find a dense domain of vectors   $\psi \in \cH_N$  so that   $H_{\La} \psi $ converges.
In  fact we show $H_{I,\La} \psi \to 0$  which suffices. 
We have $H_{I, \La} \psi = -g_{\La} B_{\La}^*  B_{\La} \psi$.   By  (\ref{sch})
$\|B^*_{\La} \|  \leq  \|\rho_{\La}\|_2N  \leq  \cO(\La)$.  Since  also  
  $g_{\La}  \to 0$ suffices to find a dense domain so  that   $\| B_{\La} \psi  \| = \cO(\La^{-1})$.

Now  $\cH_N$ can be thought of as symmetric functions  in   $\cL^2(\bbR^{2N})$.
We take the subspace of functions  in  $\cS(\bbR^{2N})$  which have a   Fourier transform 
 in $\cC^{\infty}_0 (\bbR^{2N})$ with support disjoint from 
the  hypersurfaces where points coincide.  If   $\psi$ is in this space then  
$a(p_1)a(p_2) \psi  \sim  \psi( p_1,p_2, \dots)$ is a vector-valued function 
which has a   Fourier transform in     $\cC^{\infty}_0 (\bbR^{2}\times \bbR^{2})$ 
with support disjoint  from the diagonal.   Then  
\begin{equation}
u(p,q)   \equiv   \frac{1}{\sqrt{2} (2 \pi)}    a(\frac{p}{2}+q) a(\frac {p}{2}-q) \psi 
\end{equation}
has  a Fourier transform  $\hat u(X,x)$ which is an  element 
of   $\cC^{\infty}_0 (\bbR^2 \times  ( \bbR^2- \{0\} ))$.  
Hence   $\hat v =  |x|^{-2}\hat u$ is in the same space and if $v(p,q)$ is the inverse Fourier transform    
then    $u =  -\De_q  v$.

Now we have for any  $n$     
\begin{equation}
\begin{split}
 (B_{\La} \psi)(p) 
 = & \int \rho_{\La}(q)
u(p,q)\   dq \\ 
  = & \int \rho^*_{\La}(q)
u(p,q) \  dq   + \cO(\La^{-n})\\
     =  &\int   (-\De_q  \rho^*_{\La}(q))  v(p,q)\ dq + \cO(\La^{-n}) \\
\end{split}
\end{equation}
Here we first  replace the sharp cutoff $\rho_{\La}$ by a
smooth cutoff  $\rho^*_{\La}$ and then integrate by parts.
Since    $|\De_q  \rho^*_{\La}(q)|   =  \cO(\La^{-2})$
and since   $v(p,q)$ is rapidly decreasing in both variables we  have 
 $\| B_{\La} \psi\|  =  \cO(\La^{-2})$ which suffices.

\item   This follows by the Trotter-Kato theorem.
\end{enumerate}

\bigskip

\res
\begin{enumerate}
\item  For  $N$  large our lower bound  is  $H \geq   - \mu^2e^{16\pi N^2}$. 
The coefficient $16 \pi$  can be improved but anyway the $N^2$ behavior is probably 
not optimal.  Indeed mean field 
calculations \cite{Raj99} suggest that the actual lower bound may be   $e^{\cO(N)} $.
The ground state is presumably a dense clump of  particles:  a ''bosonic star''.

\item   For further studies of the spectrum on can consider the operator 
$\Phi (E) ^{-1}$.   We note 
that  for   $E<0$ if one scales all momenta  by  $\sqrt{-E}$
the operator $\Phi (E)$ becomes 
\begin{equation}
\frac{1}{4\pi}  \log ( \frac {-E}{\mu^2} ) + W
\end{equation}
where  
\begin{equation}
\begin{split}
W   = &   (4\pi)^{-1} \int
 dp \ \chi^*(p) \log\left( H_0 + p^2/4 + 1  \right)  \chi(p)
 \\
  - & \frac{1}{2 (2 \pi)^2}   \int dp_1 dp_2  dp'_1 dp'_2 \
\chi^*(p_1 + p_2) \chi(p_1'+p_2')  \\
&\left(  a^*(p_1') a^*(p_2')\frac {1}{H_0 + \om_1  +\om_2  + \om'_1  +\om'_2+1} 
a(p_1) a(p_2) \right. \\
&\left. +\de (p_1-p_1')  a^*(p_2') \frac {4}{H_0  +\om_1  + \om_2  +\om'_2+1 }  a(p_2)
\right)
\\
\end{split}
\end{equation}
The issue is then to study properties of  $W$. 
\end{enumerate}

\subsection{extensions}

We now allow an extra inter-particle potential $v$  again assumed 
bounded.  This means we add a potential
\begin{equation}
V =  \frac 12  \int a^*(x) a^*(y)  v(x-y) a(x) a(y) \ dx dy 
\end{equation}
We have   
\begin{equation}
|(\psi, V \chi)|  \leq \frac12   \|v\|_{\infty}\| N_0^{1/2}(N_0-1)^{1/2} \psi\|\| N_0^{1/2}(N_0-1)^{1/2} \chi\|   
\end{equation}
and thus $V$ defines an operator  on  $\cH_N$ satisfying  $\| V \|   \leq N^2 \|v\|_{\infty}/2$.
This is in configuration space and we actually  consider
the momentum space version  $V' = \Ga(\cF) V \Ga (\cF^{-1})$  where    $\Gamma(\cF)$ is the induced 
Fourier transform on Fock space.
This also satisfies    $\| V' \|   \leq N^2 \|v\|_{\infty}/2$
which is the only fact we use.

With  a cutoff the full Hamiltonian is then
\begin{equation}
H^\#_{\La}  =H_0 +V' + H_{I,\La}
\end{equation}
Then   $H^\#_{\La}$ is  self-adjoint on $D(H_0)\cap  \cH_N$.
The same is true for 
\begin{equation}
H_1 = H_0 +V'
\end{equation} 
and we have     $   H_1 \geq  -N^2\|v\|_{\infty}/2$

Proceeding as before we introduce
resolvents  
\begin{equation}
R_1(E)  =(H_1-E)^{-1} \ \ \ \ \ \ \ \ \ R_{\La}^\#(E) = (H_{\La}^\#-E)^{-1}
\end{equation}
  and for   $E< -N^2 \|v\|_{\infty}/2$
\begin{equation}
\Phi^\#_{\La} (E)  =  g_{\La}^{-1}  - B_{\La}R_1(E) B_{\La}^*
\end{equation}
For such $E$ we find as in  lemma   \ref{equiv}  that $E$ is in the resolvent set of  $H^\#_{\La}$ on
$\cH_N$ if and  only if $\Phi^\#_{\La} (E)$ has a bounded inverse   on    $\tilde \cH_{N-2}$ in which case
\begin{equation}  \label{lastone} 
 R^\#_{\La}(E)   =  R_1(E)  + R_1(E) B^*_{\La}\Phi^\#_{\La} (E) ^{-1} B_{\La}R_1(E) 
\end{equation}

\begin{thm} { \ }
\begin{enumerate}
\item 
Let    $E < - e^\#_N$  where 
\begin{equation}
e_N^\#  =  \max(N,  N^2 \|v\|_{\infty},  -  \mu^2 e^{16 \pi N^2 ( C^2\|v\|_{\infty} +1)})
\end{equation}
and where  $C$ is the constant in lemma \ref{B}.
 For   $\psi \in \cH_N$
the limit  $R^\#(E)\psi   = \lim_{\La \to \infty} R^\#_{\La}(E) \psi $ 
exists and is equal to  
\begin{equation}  \label{verylastone}
 R^\#(E)   =  R_1(E)  + R_1(E) B^*\Phi^\# (E) ^{-1} B R_1(E) 
\end{equation}
\item  $R^\#(E)$ is invertible.
\item   For  $E$ complex and not in $[-e^\#_N, \infty)$ the limit  
 $R^\#(E)\psi   = \lim_{\La \to \infty} R^\#_{\La}(E) \psi $ exists.
There  is a self-adjoint operator  $H^\#(\mu)$  with  
 $H^\#(\mu)  \geq   - e^\#_N$ so that   
$R^\#(E) = (H^\#(\mu)-E)^{-1}$.
\end{enumerate}
\end{thm}

\pr  We follow the proof of theorem \ref{three}.  We have 
\begin{equation}  \label{uu}
 R_1(E)=R_0(E)-R_1(E) V' R_0(E)
\end{equation}
and hence   
\begin{equation}   \label{new}
\Phi^\#_{\La} (E)  = \Phi_{\La}(E) 
    +  B_{\La}R_0(E)V'R_0(E) B_{\La}^*      -  B_{\La}R_0(E)V'R_1(E)V'R_0(E) B_{\La}^*   
\end{equation}
For   $\La = \infty$ define  $\Phi^\#(E)$ by replacing  $\Phi_{\La}(E)$ by   $\Phi(E)$
and $B_{\La}$ by $B$.
Since   $E < - \mu^2 e^{16 \pi N^2 ( C^2\|v\|_{\infty} +1)}$ we have for  $\La$ sufficiently
large  or infinite instead of  (\ref{ee})  
\begin{equation}
\Phi_{0,\La}(E)      \geq  3 N^2 (C^2\|v\|_{\infty} + 1)
\end{equation} 
and it follows by the bound on   $\Phi_{I, \La}(E)$ that
\begin{equation}
\Phi_{\La}(E)   \geq   N^2 (C^2\|v\|_{\infty} + 1)
\end{equation}

For the  other terms in  (\ref{new})
we note  that   $E < -N$  implies    $ \| B_{\La}R_0(E) \|   \leq C $
by  lemma   \ref{B}.  Also $E <  -N^2 \|v\|_{\infty} $ and the lower
bound on   $H_1$ imply that  
  $\|R_1(E)\| \leq  ( N^2\|v\|_{\infty}/2 )^{-1}$.
Using also  $\|V'\|  \leq    N^2  \| v\|_{\infty}/2$
\begin{equation}
\begin{split}
\| B_{\La}R_0(E)V'R_0(E) B_{\La}^* \|  \leq &  C^2N^2  \| v\|_{\infty}/2 \\
\| B_{\La}R_0(E)V'R_1(E)V'R_0(E) B_{\La}^* \|  \leq &    C^2N^2 \| v\|_{\infty}/2 \\
\end{split}
\end{equation}
 Combining these  we find  for  $\La \leq  \infty$ 
\begin{equation}   \label{final}
\Phi^\#_{\La} (E)   \geq  N^2
\end{equation} 
so that   $\Phi_{\La}(E)^{-1}$ exists.
Then for  $\La < \infty$ all  $E< -e_N^\#$ are in the resolvent set for   $R^\#_{\La}(E)$
and  (\ref{lastone}) holds . 

As before   $\Phi^\#(E)$ is essentially self-adjoint  on $\tilde \cD \cap \tilde \cH_N$
On this domain    $\Phi^\#(E)\psi =  \lim_{\La \to \infty} \Phi^\#_{\La}(E)\psi$.  
This follows from the convergence for $\Phi_{\La}(E)$ and  the norm convergence
of   $B_{\La} R_0(E)$. Using the uniform bounds on the inverses
$\Phi^\#_{\La} (E)^{-1} $ converges strongly to   $\Phi^\# (E)^{-1}$.

Finally   $R^\#_{\La}(E)$  given by    (\ref{lastone})  converges strongly to    
 $R^\#_{\La}(E)$ given by  (\ref{verylastone}).   Here we use the norm convergence
of   $B_{\La} R_1(E)$ to   $B R_1(E)$ which can be demonstrated using the adjoint
of  (\ref{uu}).  This completes the proof of the first part and the second and 
third parts follow as in  theorem  \ref{three}.

\appendix

\section {Trotter-Kato Theorem}

In the text we use the following version of the Trotter-Kato theorem .

\begin{thm}  Let $\Sigma$  be a proper closed subset of  $\bbR$ and let
 $H_n$ be a sequence of self adjoint operators  with 
resolvents  $R_n(E)  = (H_n-E)^{-1}$ defined for all complex $E   \notin  \Sigma$. 
Suppose  $R_n(E)$ converges  strongly for some    $E   \notin  \Sigma$ 
and that  the limit   is invertible.  Then there exists a self-adjoint operator  $H$
with resolvents   $R(E)  = (H-E)^{-1}$ such that   $R_n(E)$ converges  strongly to   $R(E)$ 
for all complex    $E   \notin  \Sigma $.  
\end{thm}
\bigskip

A slightly different  result is proved in \cite{RS72}.   There    $\Sigma = \bbR$ 
is allowed, but one needs convergence at two points with  $\pm \textrm{Im} E >0$.   This proof can
be  easily  adapted to prove the quoted result.

\end{document}